\documentclass[preprint]{aastex}
\usepackage{wasysym}
\usepackage{apjfonts}
\usepackage{color}

\def\aj{AJ}%
\def\araa{ARA\&A}%
\def\apj{ApJ}%
\def\apjl{ApJ}%
\def\apjs{ApJS}%
\def\aap{A\&A}%
\def\mnras{MNRAS}%
\def\prc{Phys.~Rev.~C}%
\def\gca{Geochim.~Cosmochim.~Acta}%
\slugcomment{DRAFT: \today}

\shorttitle{He shell in massive stars}
\shortauthors{}

\begin{document}

\title{Silicon carbide grains of type C provide evidence
for the production of the unstable isotope $^{32}$Si in supernovae.
}
\author{M. Pignatari\altaffilmark{1,14},
E. Zinner\altaffilmark{2},
M.G. Bertolli\altaffilmark{3,14},
R. Trappitsch\altaffilmark{4,5,14},
P. Hoppe\altaffilmark{6},
T. Rauscher\altaffilmark{1,7},
C. Fryer\altaffilmark{8,14},
F. Herwig\altaffilmark{9,10,14},
R. Hirschi\altaffilmark{11,12,14},
F.X. Timmes\altaffilmark{10,13,14},
F.-K. Thielemann\altaffilmark{1}}
%M. Wiescher\altaffilmark{10}}
%F.X. Timmes\altaffilmark{3,4,10},
%R.J. de Boer\altaffilmark{2,3},
%F.-K. Thielemann\altaffilmark{1},
%C. Fryer\altaffilmark{5,10},
%A. Heger\altaffilmark{6,10},
%F. Herwig\altaffilmark{3,7,10},
%R. Hirschi\altaffilmark{8,9,10}}
\altaffiltext{1}{Department of Physics, University of Basel, Klingelbergstrasse 82, CH-4056 Basel, Switzerland}
\altaffiltext{2}{Laboratory for Space Sciences and the Physics Department, Washington University St.
Louis, MO 63130, USA}
\altaffiltext{3}{Theoretical Division (T-2), LANL, Los Alamos, NM, 87545, USA.}
\altaffiltext{4}{Department of the Geophysical Sciences, University of Chicago, Chicago, IL 60637, USA.}
\altaffiltext{5}{Chicago Center for Cosmochemistry}
\altaffiltext{6}{Max Planck Institute for Chemistry, Hahn-Meitner-Weg 1, 55128 Mainz, Germany.}
\altaffiltext{7}{Centre for Astrophysics Research, School of Physics, Astronomy
and Mathematics, University of Hertfordshire, College Lane, Hatfield AL10
9AB, UK.}
\altaffiltext{8}{Computational Physics and Methods (CCS-2), LANL, Los Alamos, NM, 87545, USA.}
\altaffiltext{9}{Department of Physics \& Astronomy, University of Victoria, Victoria, BC, Canada.}
\altaffiltext{10}{The Joint Institute for Nuclear Astrophysics, Notre Dame, IN 46556, USA}
\altaffiltext{11}{Keele University, Keele, Staffordshire ST5 5BG, UK.}
%\altaffiltext{12}{Institute for the Physics and Mathematics of the Universe, University of Tokyo, 5-1-5 Kashiwanoha, Kashiwa 277-8583, Japan}
\altaffiltext{12}{Kavli Institute for the Physics and Mathematics of the Universe (WPI), University of Tokyo, 5-1-5 Kashiwanoha, Kashiwa, 277-8583, Japan}
\altaffiltext{13}{Arizona State University, School of Earth and Space Exploration, PO Box 871404, Tempe, AZ, 85287-1404, USA.}
%\altaffiltext{10}{The Joint Institute for Nuclear Astrophysics, Notre Dame, IN 46556, USA}
%\altaffiltext{4}{Arizona State University, School of Earth and Space Exploration, PO Box 871404, Tempe, AZ, 85287-1404, USA.}
%\altaffiltext{5}{Computational Physics and Methods (CCS-2), LANL, Los Alamos, NM, 87545, USA.}
%\altaffiltext{6}{Monash Centre for Astrophysics, School of Mathematical Sciences, Monash University, Vic 3800, Australia.}
%\altaffiltext{7}{Department of Physics \& Astronomy, University of Victoria, Victoria, BC,
%\altaffiltext{8}{Keele University, Keele, Staffordshire ST5 5BG, UK.}
%\altaffiltext{9}{Institute for the Physics and Mathematics of the Universe, University of Tokyo, 5-1-5 Kashiwanoha, Kashiwa 277-8583, Japan}
%V8P5C2 Canada.}
%\altaffiltext{3}{Dipartimento di Fisica Generale, Universit\'a di Torino,
%Via Pietro Giuria 1, Torino 10125, Italy; gallino@ph.unito.it}
\altaffiltext{14}{NuGrid collaboration,  \url{http://www.nugridstars.org}}
%\altaffiltext{5}{Geneva Observatory, CH-1290 Sauverny, Switzerland;
%georges.meynet@obs.unige.ch}
%\altaffiltext{6}{IPMU, University of Tokyo, Kashiwa, Chiba 277-8582, Japan.}
%\altaffiltext{7}{Department of Physics \& Astronomy, University of Victoria, Victoria, BC,
%V8P5C2 Canada; fherwig@uvic.ca}

%\include{abstract}
\begin{abstract}
Carbon-rich grains are observed to condense in the ejecta of recent
core-collapse supernovae, within a year after the explosion.  Silicon
carbide grains of type X are C-rich grains with isotpic signatures of
explosive supernova nucleosynthesis have been found in primitive
meteorites.  Much rarer silicon carbide grains of type C are a
special sub-group of SiC grains from supernovae.  They show peculiar
abundance signatures for Si and S, isotopically heavy Si and
isotopically light S, which appear to to be in disagreement with model
predictions.  We propose that C grains are formed mostly from C-rich
stellar material exposed to lower SN shock temperatures than the more
common type X grains.  In this scenario, extreme $^{32}$S enrichments
observed in C grains may be explained by the presence of short-lived
$^{32}$Si ($\tau$$_{1/2}$ = 153 years) in the ejecta, produced by
neutron capture processes starting from the stable Si isotopes.  No
mixing from deeper Si-rich material and/or fractionation of Si from S
due to molecular chemistry is needed to explain the $^{32}$S
enrichments.  The abundance of $^{32}$Si in the grains can provide
constraints on the neutron density reached during the supernova
explosion in the C-rich He shell material. The impact of the large
uncertainty of the neutron capture cross sections in the $^{32}$Si
region is discussed.
%sets a limit
%on the accuracy of predictions of the $^{32}$Si/$^{28}$Si ratio.
\end{abstract}

%\keywords{stars: abundances --- chemically peculiar --- early-type --- rotation }

\section{Introduction}
\label{sec:intro}

Despite recent improvements in simulations of core-collapse supernova
(CCSN) explosions \citep[e.g.,][]{janka:12} the understanding of
supernova still has major gaps, and observations of SN and their
ejecta still provide many puzzles \citep[e.g.,][and references
therein]{fryer:12}. Of particular importance may be the assymmetric
nature of the explosion and the hydrodynamic development of the layers
ejected after the explosion
\citep[e.g.,][]{kiaer:10,isensee:10,delaney:10}.

Several types of pre-solar grains from primitive carbonaceous
meteorites that are associated with SN nucleosynthesis due to their
isotopic ratios \citep[see e.g.,][]{clayton:04,zinner:13} provide
constraints on these explosions.  Pre-solar grains carry the
signatures of their stellar origin, and their interpretation may help
to guide CCSN models.

Silicon carbide is one of the types of stardust grains that have
been identified in primitive meteorites \citep[e.g.,][]{zinner:13}.
While most of these so-called presolar SiC grains originate
in Asymptotic Giant Branch stars, there are two rare sub-types of
SiC grains that have a CCSN origin.
Type X grains (about 1\% of all presolar SiC grains),
have large excesses in $^{28}$Si. This signature and evidence for the
initial presence of $^{44}$Ti in a subset of these grains is proof of
their SN origin: both isotopes are predicted to be abundant in the Si/S zone
of supernovae \citep[][]{rauscher:02}.
More recently, \cite{pignatari:13a}, hereafter P13, showed that
$^{28}$Si and $^{44}$Ti may also be produced at the bottom of the He shell
exposed to high shock velocities and/or high energies,
reproducing several isotopic abundance patterns typical of SiC X grains
and graphites from SNe.

Silicon carbide grains of type C are even rarer (about 0.1\% of all SiC grains)
than SiC X grains.
They have a large excess in $^{29}$Si and $^{30}$Si and most of them have been
found by automatic searches in the  NanoSIMS detection apparatus.
%In addition,
Some of these grains contain extinct $^{44}$Ti, similar to SiC-X grains.
Just over a dozen of these grains have been
identified, and
%\citep[][]{amari:99,croat:10,gyngard:10,hoppe:10,zinner:10,hoppe:12,
%orthous-daunay:12,xu:12}.
%(Amari, Zinner, & Lewis 1999; Croat, Stadermann, & Bernatowicz 2010; Gyngard,
%Nittler, & Zinner 2010; Hoppe et al. 2010; Hoppe, Fujiya, & Zinner 2012; Zinner, Gyngard,
%& Nittler 2010; Orthous-Daunay et al. 2012; Xu et al. 2012).
nine have been analyzed for their S isotopic ratios, showing
large $^{32}$S excesses, with $^{32}$S/$^{33,34}$S ratios ranging up
to 16 times solar
\citep[][]{amari:99,croat:10,gyngard:10,hoppe:10,hoppe:12,zinner:10,
orthous-daunay:12,xu:12}.
%\citep[][]{gyngard:10,hoppe:10,hoppe:12,orthous-daunay:12,xu:12}.
This is puzzling, because in existing SN models the only zone with
large $^{32}$S excesses
is the Si/S zone \citep[][]{meyer:95}, which has large $^{28}$Si excesses,
whereas zones with $^{28}$Si depletions (i.e., $^{29,30}$Si
excesses) are predicted to have also $^{32}$S depletions
\citep[e.g.,][]{rauscher:02}. \cite{hoppe:12}
have invoked element fractionation between sulfur and silicon by molecule chemistry in
the SN ejecta to explain this result. However, this ad hoc explanation
%leaves us uncomfortable, especially in view of the S isotopic composition of
cannot explain all the data,
especially the S isotopic composition of
one C grain with $\delta$($^{33}$S/$^{32}$S) and $\delta$($^{34}$S/$^{32}$S)
values being as low as -940\permil~\citep[][]{xu:12},
even more extreme than those of S in the Si/S zone.
In this paper we propose that the $^{32}$S excesses in C grains
are due to the radioactive decay of short-lived $^{32}$Si
\citep[$\tau_{1/2} = 153$ years,][]{ouellet:11}.
%($\tau_{1/2} = 153$
%\footnote{http://www.nndc.bnl.gov/nudat2/indx_dec.jsp}
%years).
We present models of explosive nucleosynthesis in the inner part of the
He/C zone where $^{29}$Si and $^{30}$Si as well as $^{32}$S excesses can
be produced while
maintaining a C-rich environment.

The paper is organized as follows. In \S \ref{sec: models_description}
we describe the stellar models and the nucleosynthesis calculations,
in \S \ref{sec: comparison} we compare theoretical results with
measurements for C grains.
Finally, in \S \ref{sec: summary} we give our conclusions.
%results are summarized.

%\include{stellar_calculations}
\section{Stellar model calculations and nucleosynthesis}
\label{sec: models_description}

This investigation is based on seven SN explosion models for a
$15~\mathit{M_{\odot}}$, $\mathit{Z} = 0.02$ star, three of which were
introduced in P13.  The pre-supernova evolution is calculated with the
code GENEC \citep[][]{eggenberger:08}.  The explosion simulations
include the fallback prescription by \cite{fryer:12}, and are
performed for a case with recommended initial shock velocity and six
cases where the latter is reduced by a factor of 2, 4, 5, 10, 20 and
100, respectively (models 15r, 15r2, 15r4, 15r5, 15r10, 15r20 and
15r100).  The standard initial shock velocity used beyond fallback is
$2\times10^{9} \mathrm{cm ~s}^{-1}$.
%In a one-dimensional CCSN explosion model,
The kinetic explosion
energy for these 15 M$_{\odot}$ models ranges from
$4-5\times10^{51} \mathrm{ergs}$ to less than $10^{51} \mathrm{ergs}$.
The post-processing code MPPNP is used
to calculate the nucleosynthesis in the star before and
during the explosion \citep[see e.g.,][]{bennett:12}.
%\citep[see][]{pignatari:13}.
For the present study we focus only on the C-rich
explosive He burning layers, including the He/C zone and
a small part of the O/C zone.

\begin{figure}
\centering
\resizebox{10.5cm}{!}{\rotatebox{0}{\includegraphics{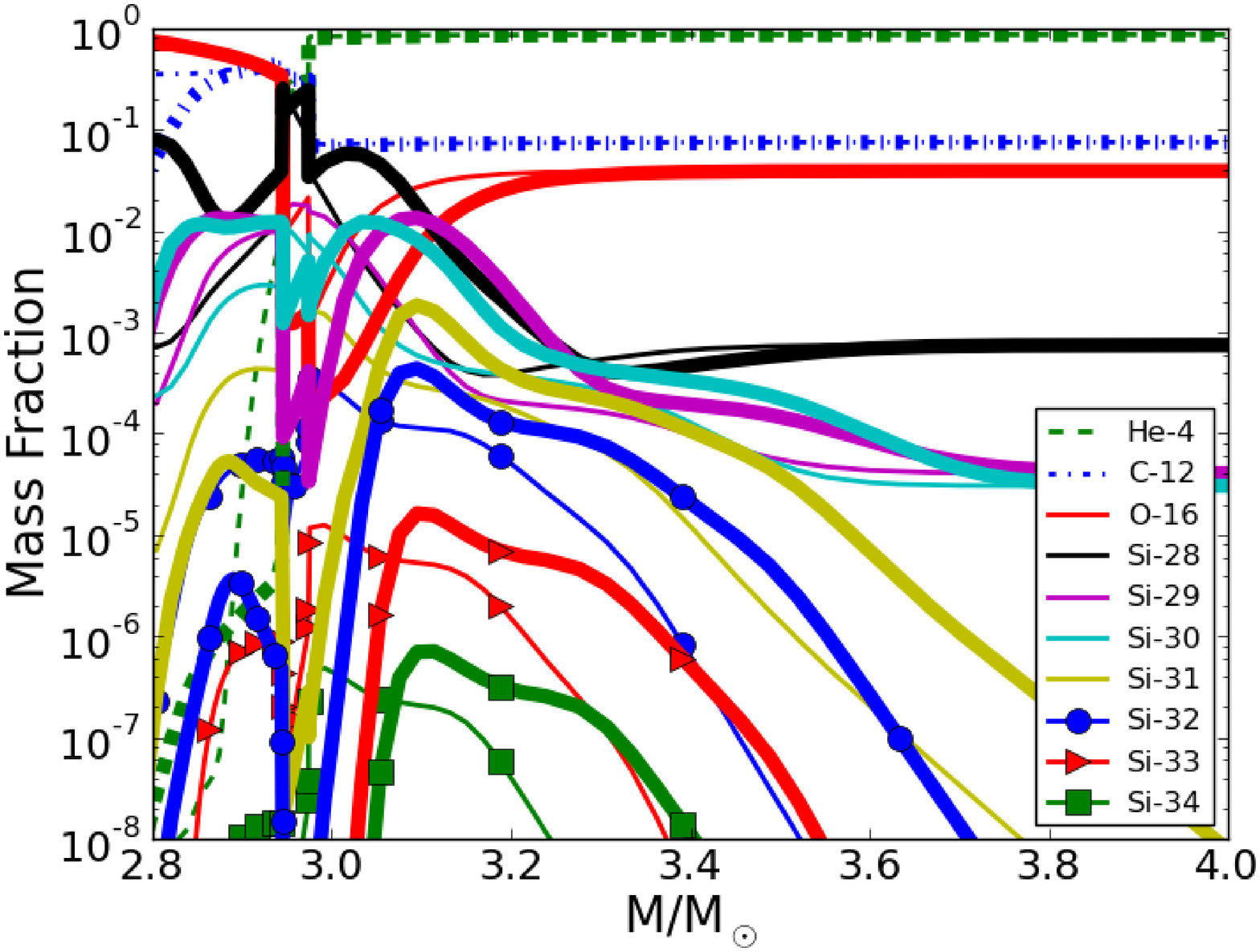}}}
\caption{
Isotopic abundance profiles given 2.5 seconds after the explosion
in the top of the O/C zone, the C/Si zone, and the He/C zone of the
15 M$_{\odot}$ SN models 15r and 15r4.
Shown are profiles for $^4$He, $^{12}$C, $^{16}$O,
and the Si isotopes along the neutron capture chain from $^{28}$Si
to $^{34}$Si.
The models 15r and 15r4 are represented by
thick and thin lines, respectively.
}
\label{fig:si_chain}
\end{figure}

The abundances of key species and $^{28-34}$Si
are reported in Fig.\,\ref{fig:si_chain} for models 15r and 15r4.
Results are similar for the intermediate model 15r2.
The bottom of the He/C zone is strongly affected by the explosion.
While $^{12}$C is not significantly modified, $^{16}$O is depleted
and feeds the production of heavier $\alpha$-isotopes, including
$^{28}$Si. This stellar region was defined as the C/Si zone
in P13. The main reason for this behavior is the higher
$\alpha$-capture rates starting from the $^{16}$O($\alpha$,$\gamma$)$^{20}$Ne reaction
than that of the $^{12}$C($\alpha$,$\gamma$)$^{16}$O reaction at explosive
He shell temperatures (as explained by P13).
Models with lower shock velocities show weaker explosion signatures. In particular,
model 15r100 does not show any significant departures from pre-explosive abundances
during the explosion in the C-rich region.

\begin{figure}
\centering
\resizebox{11.5cm}{!}{\rotatebox{0}{\includegraphics{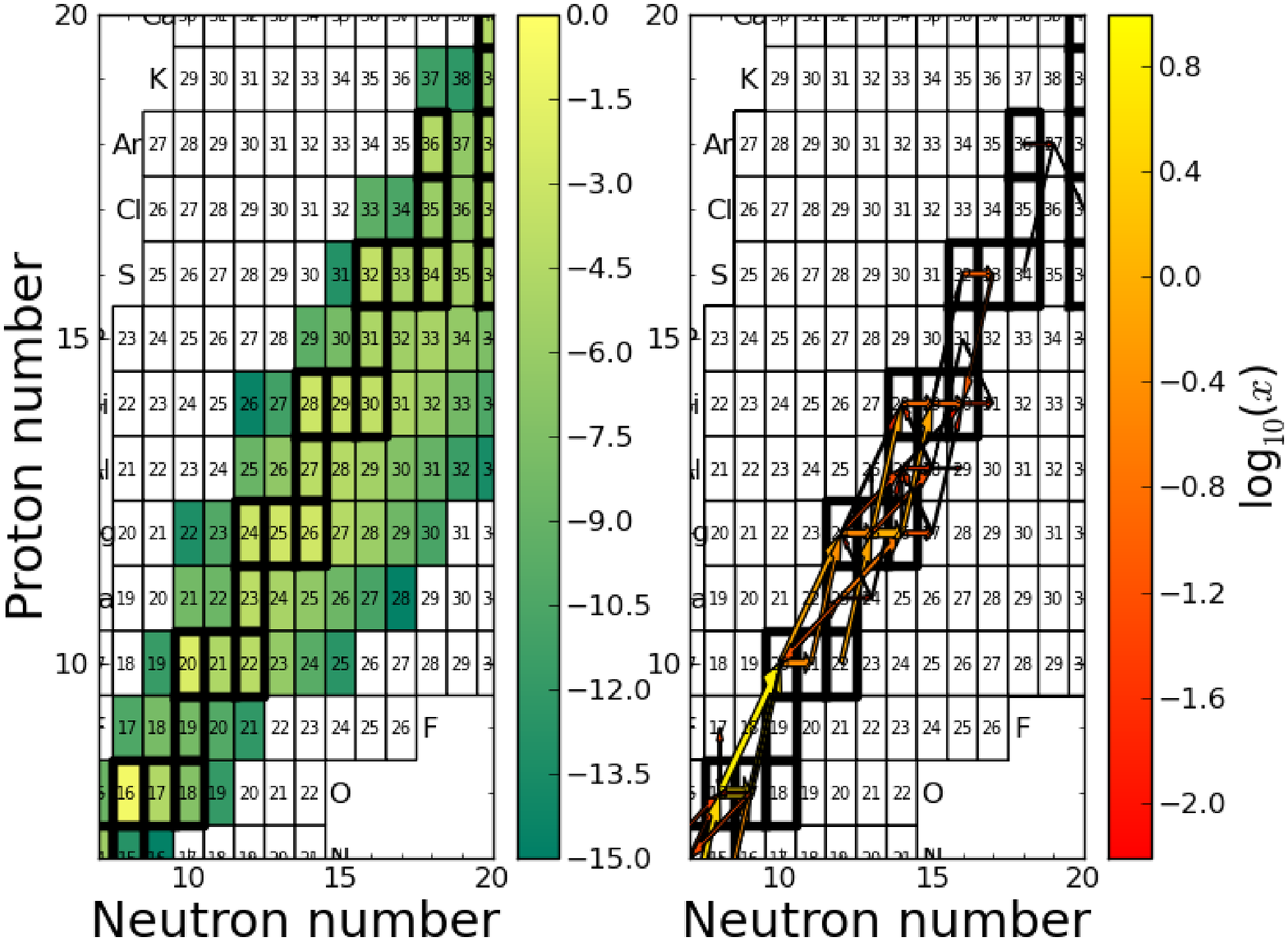}}}
\resizebox{11.5cm}{!}{\rotatebox{0}{\includegraphics{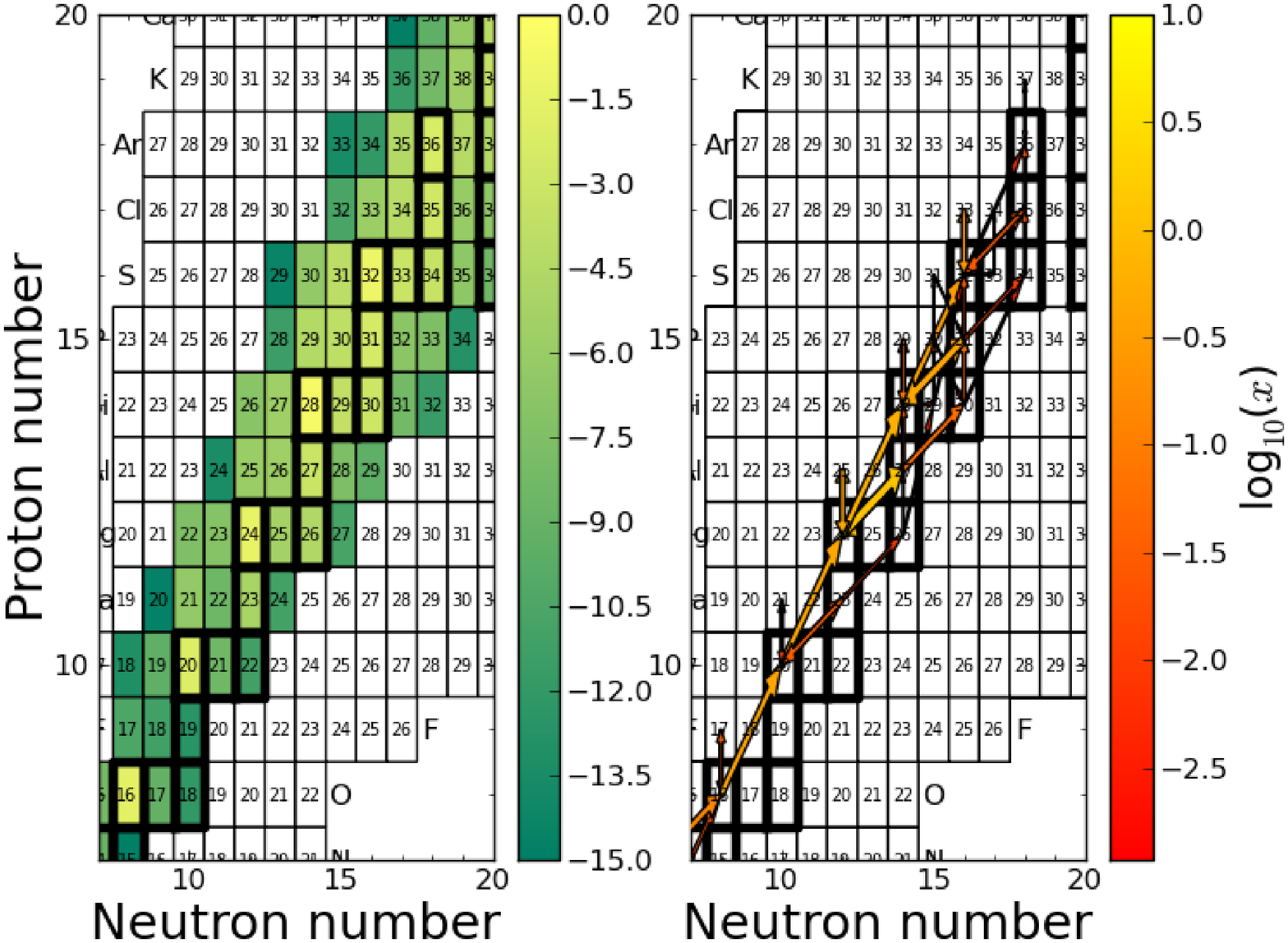}}}
\caption{
The mass fraction abundance distributions (green, left) and nucleosynthesis
fluxes (arrows with red to yellow color, right)
at $\sim 10^{-5} \mathrm{sec}$ (upper panels) and $\sim 10^{-2} \mathrm{sec}$ (lower panels)
after the explosion
%temperature peak (T $\approx 2\times10^{9} \mathrm{K}$) is reached
at mass coordinate 2.95 M$_{\odot}$ of the model 15r.
%Abundances are given in mass fractions.
The nucleosynthesis fluxes,
[$\delta$$Y_{\rm i}$/$\delta$t]$_{\rm j}$, show the variation of the
abundance $Y_{\rm i}$ = $X_{\rm i}$/$A_{\rm i}$ due to the reaction j.
The arrow width and color correspond to the flux strength.
Heavy-lined boxes correspond to the stable isotopes.
}
\label{fig:abund_and_flux}
\end{figure}

Along the Si neutron capture chain, $^{29-30}$Si and heavier unstable
Si species are produced efficiently by neutron captures starting from $^{28}$Si.
The larger explosion temperatures in model 15r than in model 15r4
are pushing the production peaks of different
Si neutron-rich species to larger mass coordinates,
not significantly affecting their absolute abundance.
Therefore, abundance yields for the Si isotopes in the explosive He shell
result from the interplay between $\alpha$-captures
and neutron captures, triggered by activation of the $^{22}$Ne($\alpha$,n)$^{25}$Mg
neutron source \citep[e.g.,][and references therein]{meyer:00}.
The main abundance features and dominating nucleosynthesis fluxes
for two different times of the SN explosion are given
in Fig.\,\ref{fig:abund_and_flux}, in the so-called C/Si zone
(M $\sim$ 2.95 M$_{\odot}$, model 15r, see also P13).
In the early stages of the explosion, depending on the available $^{22}$Ne,
the $\alpha$-capture path starting from $^{16}$O is accompanied
by (n,$\gamma$)($\alpha$,n) sequences, producing the same $\alpha$-species.
An example is $^{20}$Ne($\alpha$,$\gamma$)$^{24}$Mg and
$^{20}$Ne(n,$\gamma$)$^{21}$Ne($\alpha$,n)$^{24}$Mg.
During the later stages of the explosion and/or low $^{22}$Ne abundances, the
($\alpha$,$\gamma$) fluxes become dominant. Note that for explosive
He burning conditions the ($\alpha$,p) fluxes are compensated
by their reverse reactions, and proton captures on the abundant
$\alpha$-isotopes do not affect the abundance of their parent species
because of the efficient reverse ($\gamma$,p) photodisintegrations.
%Neutron-rich Si species are mostly produced in regions
%of the He shell above the C/Si zone (see Fig.\,\ref{fig:si_chain}).
%Indeed, the explosion temperature decreases as one moves outward in the star,
%and ($\gamma$,n)-photodisintegrations become less efficient in compensating the
%neutron capture flux.

\begin{figure}
\resizebox{11.5cm}{!}{\rotatebox{0}{\includegraphics{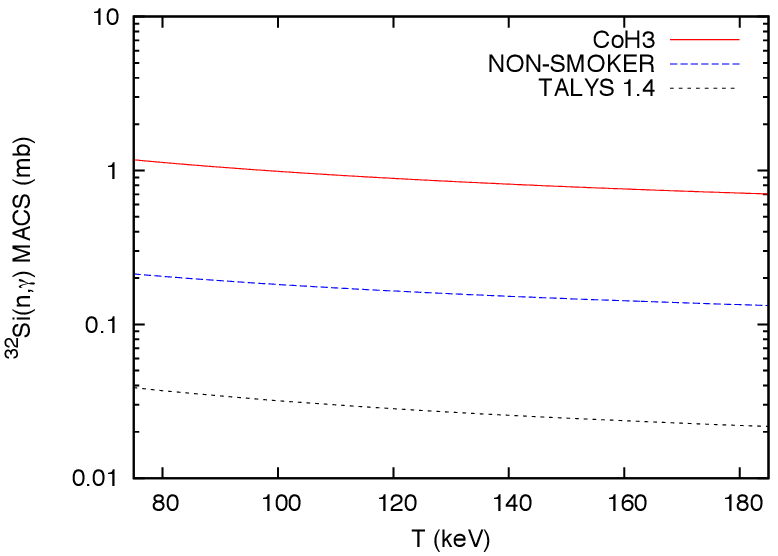}}}
\resizebox{11.5cm}{!}{\rotatebox{0}{\includegraphics{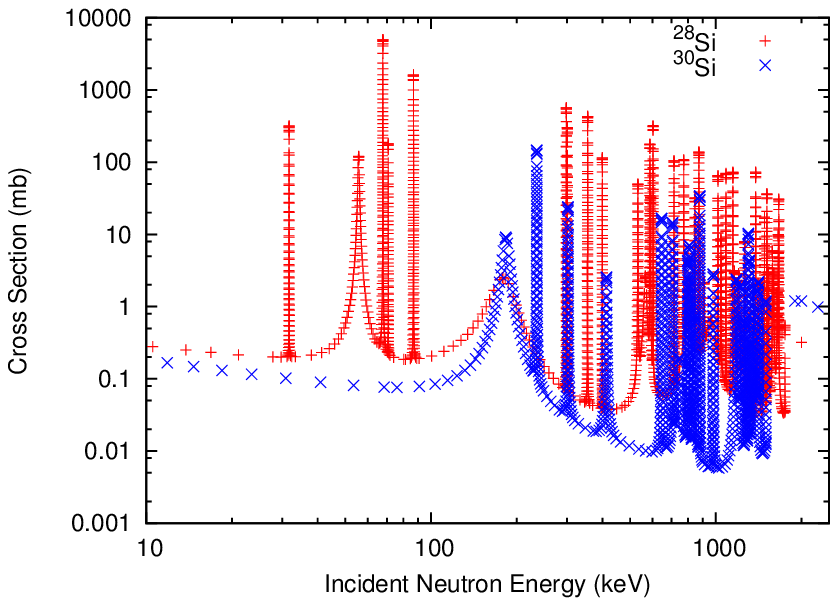}}}
\caption{
Upper panel: neutron capture MACS for $^{32}$Si, calculated by
different statistical HF models in the temperature range of interest
$T=1-2\times10^9 \mathrm{K}$ (corresponding to $\sim 90-170$ keV).
No experimental data exist for $^{32}$Si,
therefore one has to rely on theoretical calculations.
Statistical methods are not applicable in the primary energy range of interest
for this nucleus, therefore a more appropriate approach is needed to constrain the uncertainty.
Lower panel: neutron capture cross-sections for $^{28}$Si and $^{30}$Si
from the ENDF/B-VII.1 library.
 (see end of \S \ref{sec: models_description} for discussion).%The level density indicated
%by the resonances is sufficiently high only above $\sim 600$ keV for statistical HF calculations
%to be applicable.
%Upper panel:  $^{32}\mathrm{Si}(\mathrm{n},\gamma)$ MACS according to HF codes {\tt CoH$_3$}~\citep[][]{Kawano2004}
%and {\tt TALYS 1.4}~\citep[][]{Koning2008,Koning2011},
%and {\tt NON-SMOKER}.
%Lower panel: n-capture cross-sections for $^{28}$Si and $^{30}$Si
%from the ENDF/B-VII.1 library (see end of \S \ref{sec:
%  models_description} for discussion).
%%At a stellar temperature of $T=10^9 \mathrm{K}$,
%%these resonances are included in a MACS.
}
\label{fig:32SiHF_and_SiResonances}
\end{figure}

In the present calculations, we use for the (n,$\gamma$)
reactions on unstable Si species
the rates from Hauser-Feshbach (HF) calculations
by the {\tt NON-SMOKER} code \citep[][]{rauscher:00},
available in JINA REACLIB
v1.1~\citep[e.g.,][]{cyburt:10}.
The uncertainties of the neutron capture rates
in the mass region of $^{32}$Si are very large.
Figure~\ref{fig:32SiHF_and_SiResonances}, upper panel,
shows the Maxwellian averaged cross section (MACS)
for neutron capture on $^{32}$Si as calculated from
the HF codes {\tt CoH$_3$}~\citep[][]{Kawano2004}
and {\tt TALYS 1.4}~\citep[][]{Koning2008,Koning2011},
and {\tt NON-SMOKER}.
At a temperature near 90 keV ($\sim10^9  \mathrm{K}$) we see a difference
of almost two orders of magnitude between the highest
(from {\tt CoH$_3$}) and the lowest calculated values ({\tt TALYS 1.4}).
%This provides an estimate of a factor of 100 on the lower bound of the uncertainty
%of the $^{32}$Si neutron capture MACS.
Notice, however, that these theoretical predictions are still consistent
within the
large uncertainty of the $^{32}$Si(n,$\gamma$)$^{33}$Si rate.
The large uncertainty is due to the nuclear level
density being too low to apply the HF model for neutron-rich isotopes of
Si. The model relies on the statistical averaging over levels in the compound
nucleus and thus a sufficiently high nuclear level density is required
at the compound formation energy~\citep[][]{rauscher:97}.
The ENDF/B-VII.1 library of~\cite{Chadwick2011} provides the
location of neutron capture resonances for even-even nuclei near $^{32}$Si,
shown in Fig.~\ref{fig:32SiHF_and_SiResonances}, lower panel.
While no data are available on $^{32}$Si neutron capture resonances,
the neighboring even-even nuclei $^{30}$Si and $^{28}$Si give an indication of the
number of levels accessible at different incident particle energies.  Above an energy
of $\sim 600$ keV statistical methods become appropriate.
For this reason, we considered an uncertainty of a factor of 100
for the $^{32}$Si neutron capture cross section. The impact of this uncertainty
is presented in \S \ref{sec: comparison}.

Where experimental knowledge of the single resonances has been obtained, such as in the
case of $^{28}$Si and $^{30}$Si, uncertainties may still arise from the precise
location and strength of each resonance.  However, uncertainties from experiment
are expected to be much lower than those introduced by the use of HF calculations in an
inappropriate region.

\section{Comparison with observations}
\label{sec: comparison}

\begin{figure}
\resizebox{11.5cm}{!}{\rotatebox{0}{\includegraphics{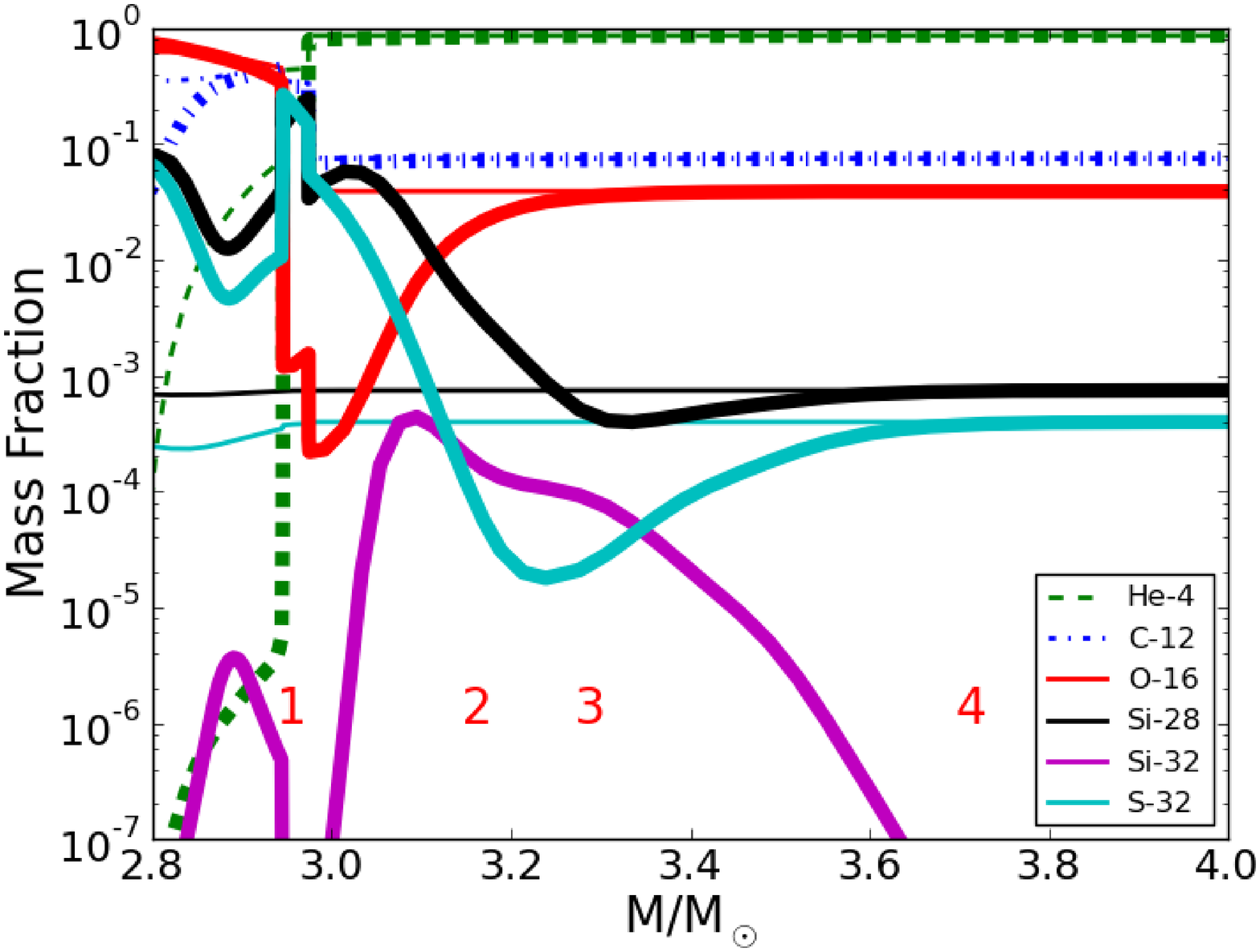}}}
\resizebox{11.5cm}{!}{\rotatebox{0}{\includegraphics{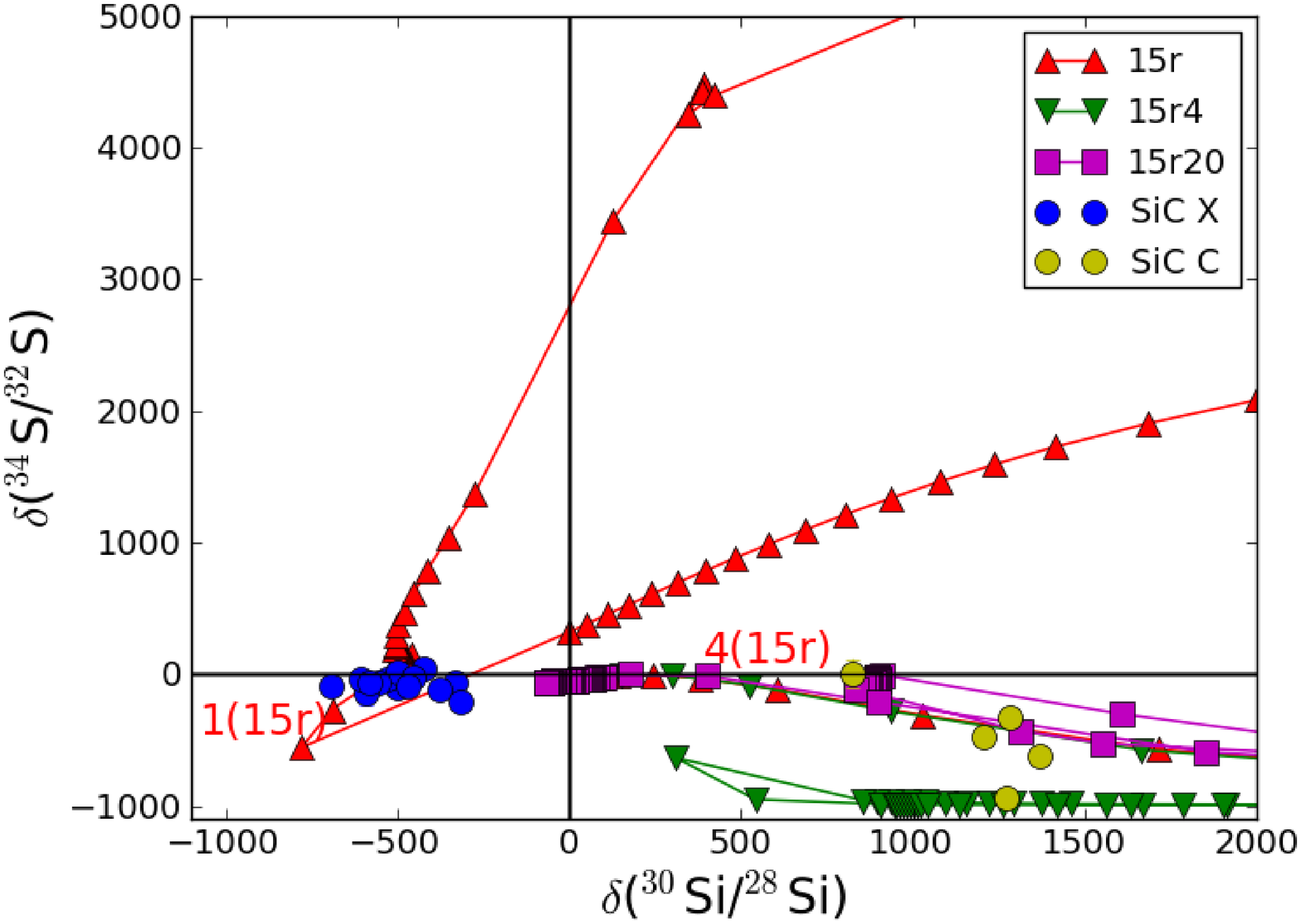}}}
\caption{
Upper panel: Abundances of selected isotopes in the outer part
of the O/C zone and the He/C zone for model 15r before and
after CCSN explosion (thin and thick lines).
Lower panel: The $^{30}$Si/$^{28}$Si and $^{34}$S/$^{32}$S
isotopic ratios of presolar SiC X and C grains, plotted
as $\delta$-values, deviations from the solar ratios in parts
per thousand (\permil), are compared with predictions of
three different models (15r, 15r4 and 15r20) in the mass range
shown in the upper panel.
We highlight the predicted Si and S isotopic ratios
of model 15r at two different mass coordinates,
M = 2.95 and 3.7 M$_{\odot}$ (see upper panel).
The Si isotopic ratios of zones "2" and "3" are located
out of the plot range, with
$\delta$($^{30}$Si) $\sim$ 15000 and $\delta$($^{34}$S) $\sim$ $-$1000.
A fractionation factor of 10$^4$ for Si/S is assumed.
%Notice that
%likely all S in SiC-C is due to contamination. In the plot, "SiC C" grains refer to multiple
%references in \cite{hynes:09}, "SiC C" grains to \cite{hoppe:12}.
}
\label{fig:si_s_delta_grains}
\end{figure}

We compare in Fig.\,\ref{fig:si_s_delta_grains} the abundances from the C-rich
ejecta  from our models (\S \ref{sec: models_description}) originating
from the C/Si zone,
the whole He/C zone and the C-rich part of the He/N zone,
with isotopic ratios of single SiC X and C grains
%\citep[][INCLUDE HERE ALL THE REFERENCES NEEDED]{hoppe:12}.
from the St.\, Louis Presolar Grains Database
\citep[][]{hynes:09}.
No mixing between layers is considered and
%Mildly C-rich regions in the
%He/N zone (H-burning layer) will be discussed in a future work.
 SiC-X and C grains with $^{12}$C/$^{13}$C lower than solar are excluded.
They are not reproduced by these models that have
high C isotopic ratio.
%Furthermore, since in the present models the C-rich material
%in the He shell show high $^{12}$C/$^{13}$C larger than solar,
%we will exclude from the direct comparison the grains
%with low C ratio. Also for them $^{32}$Si is likely
%the source of the $^{32}$S enrichment. SiC-X and C grains
%with $^{12}$C/$^{13}$C lower than solar need to be considered separately,
%and will be discussed elsewhere.

The standard model (15r, upper panel, layer 1
Fig.\,\ref{fig:si_s_delta_grains}) shows a strong $^{28}$Si and
$^{32}$S production and absence of $^{32}$Si in the C/Si zone during
the explosion (see also P13).  Outward, in the inner part of the He/C
zone, the lower explosion temperatures and the neutron burst triggered
by the $^{22}$Ne($\alpha$,n)$^{25}$Mg \citep[$n$ process,
e.g.,][]{meyer:00} gradually reduce $^{28}$Si- and
$^{32}$S-enrichments, whereas $^{32}$Si is synthesized and accumulated
according to its neutron capture cross section (as discussed in \S
\ref{sec: models_description}).  The outer parts of the He/C zone show
mild enrichments of the stable neutron-rich Si and S isotopes due to
pre-explosive $s$-processing.

The $^{28}$Si-excess observed in SiC X grains
are reproduced in parts of the C/Si zone
for the models 15r and 15r2 (e.g, layer $1$ of model 15r, Fig.\,\ref{fig:si_s_delta_grains},
lower panel). SiC C grains show larger $^{32}$S-excesses than SiC X
grains, and positive $\delta$($^{30}$Si).
Such a signature is consistent with abundance predictions from
more external zones in the C-rich He shell.
In models 15r4-15r20 the shock temperature is not sufficient to reproduce
the $^{28}$Si-excess observed in SiC X grains (see also P13).
Conversely, the presented models can reproduce the Si and S isotopic
ratios in the C grains over a large range in initial shock
velocities.
Also in case of contamination or mixing with isotopically
more normal material (see P13),
the grain signatures can be explained since $\delta$($^{30}$Si) values
up to $\sim$ 15000$-$20000 (e.g., models 15r, 15r2 and 15r4, zones "2" and "3")
are associated with large $^{32}$S enrichments ($\delta$($^{34}$S) $\sim$ $-$1000,
 Fig.\,\ref{fig:si_s_delta_grains}, lower panel, outside the plot range).
For most of the He shell material the $^{32}$Si signature
dominates S isotopic anomalies, assuming an arbitrary
Si/S fractionation of 10$^4$ during grain formation
(Fig.\,\ref{fig:si_s_delta_grains}, lower panel).
This assumption expresses the hypothesis
that all $^{32}$S observed in C grains originates from the decay of $^{32}$Si
(see below for details).
Only little S condenses into SiC grains, justifying the assumed
elemental fractionation \citep[e.g.,][]{amari:95}.

\begin{figure}
\resizebox{11.5cm}{!}{\rotatebox{0}{\includegraphics{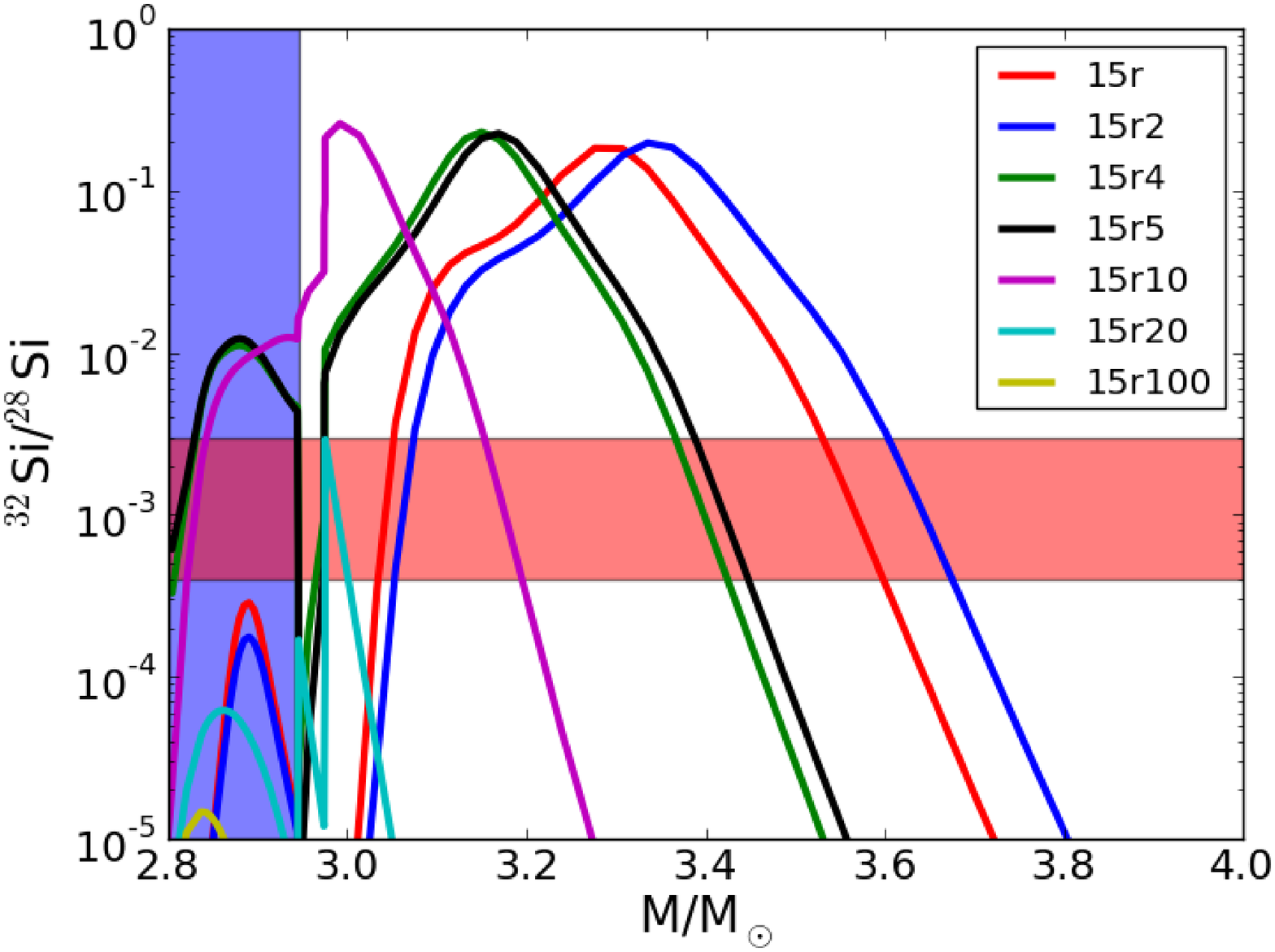}}}
\resizebox{11.5cm}{!}{\rotatebox{0}{\includegraphics{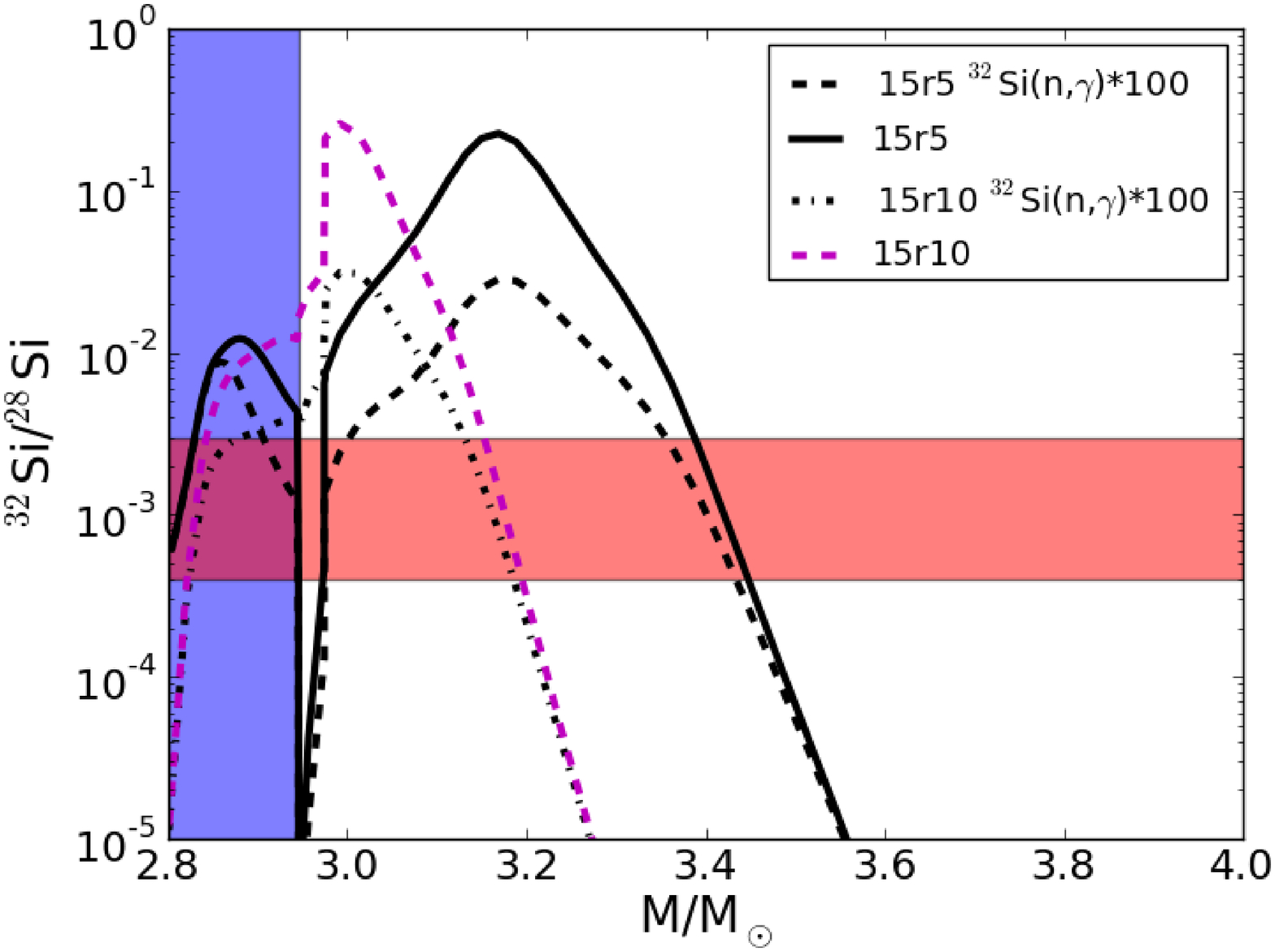}}}
\caption{
Upper panel: Final isotopic ratio $^{32}$Si/$^{28}$Si in the C-rich
explosive He shell for models with initial shock velocities
varying by a factor of 100 (decreasing from model 15r to 15r100).
The blue-shaded area denotes O-rich material.
For comparison, the red-shaded area indicates the range
of $^{32}$Si/$^{28}$Si ratios, inferred from S isotopic ratios and
S and Si abundances in presolar SiC C grains, under
the assumption that the measured $^{32}$S excess derives
from $^{32}$Si decay.
Lower panel: Impact on the $^{32}$Si/$^{28}$Si ratio from increasing
the $^{32}$Si neutron capture cross section by a factor of 100
for models 15r5 and 15r10.
}
\label{fig:si32dsi28}
\end{figure}

In Fig.~\ref{fig:si32dsi28}, upper panel, we show the $^{32}$Si/$^{28}$Si isotopic ratios
from different models described in \S \ref{sec: models_description}, comparing them with
the ratios inferred for C grains from the radiogenic
$^{32}$S.
We estimated the ratio of the radioactive $^{32}$Si ($^{32}$S$^*$) to $^{28}$Si
and thus the original $^{32}$Si/$^{28}$Si ratio by
assuming that all the S (S$_{\rm tot}$)
in the grains was either $^{32}$S$^*$ or
isotopically normal S (S$_{\rm norm}$) from contamination.
The latter assumption is based on the fact that S is volatile
and is not likely to condense into SiC. The grains are
therefore expected to contain only marginal intrinsic S.
Second, the S concentrations are low
in the He shell layers with no $^{28}$Si enrichment.
%in the SN region we consider
%in the stellar models.
Finally, some of the S isotopic images of the C grains
measured showed $^{34}$S to be more abundant at the edges of the grains
and $^{32}$S excesses to be higher in interior than in border
regions.
%From the measured ion ratios $^{32}$S$^-$/$^{28}$Si$^-$
We determined
the atomic $^{32}$Si/$^{28}$Si ratios by applying a S$^-$/Si$^-$
sensitivity factor of 3, inferred from measurements of
Si and S ion yields on synthetic SiC and Mundrabilla FeS,
respectively \citep[][]{hoppe:12}.
Since $^{32}$S$_{\rm tot}$ = $^{32}$S$^*$+$^{32}$S$_{\rm norm}$ and
$^{32}$S$^*$ = $-0.001\times\delta$S$\times$($^{32}$S$^*$+$^{32}$S$_{\rm norm}$),
we obtained the $^{32}$S$^*$/$^{28}$Si ratios by multiplying
$^{32}$S/$^{28}$Si with $-0.001\times\delta$S.
Here $^{32}$S$_{\rm norm}$ is $^{32}$S of the isotopically
normal component S$_{\rm norm}$
(assumed to be contamination). For $\delta$S
we took the average of $\delta$($^{33}$S/$^{32}$S) and
$\delta$($^{34}$S/$^{32}$S).
Within errors the latter two values are equal for all measured grains,
providing additional evidence that we are dealing just with an excess in
$^{32}$Si.
In Fig.~\ref{fig:si32dsi28}, we show that the observed range of
$^{32}$Si/$^{28}$Si ratios is matched by predictions
from stellar models at different energies, in agreement with
Fig.\,\ref{fig:si_s_delta_grains}.
%In general, with increasing
%explosion temperatures the observed isotopic ratios are reproduced
%at increasing interior masses in the He shell.
Typical conditions required for matching the inferred
$^{32}$Si/$^{28}$Si ratios (e.g., at M = 3.4 M$_{\odot}$ for models 15r4 and 15r5)
have a peak temperature of $\sim 8\times10^{8} \mathrm{K}$ and a neutron density peak
of $\sim 10^{18-19} \mathrm{cm}^{-3}$, with a $^{28}$Si mass fraction of
$\sim 5\times10^{-4}$. The models 15r-15r5 with the highest explosion temperatures
also fit the observed $^{32}$Si/$^{28}$Si ratio deeper in the He shell
(e.g., at M = 3.05 M$_{\odot}$ for model 15r and 15r2), with a
$^{28}$Si mass fraction of $\sim 5\times10^{-2}$. In these cases,
the temperature peak is about $1.6\times10^{9} \mathrm{K}$, with a neutron density
peak of a few $10^{22} \mathrm{cm}^{-3}$ for few 10$^{-5} \mathrm{sec}$, dropping
quickly to densities more typical of the $n$ process.

Since the grains may contain some
normal component (P13), the inferred $^{32}$Si/$^{28}$Si
needs to be considered a lower limit of the original ratio
in the He shell material.
In Fig.~\ref{fig:si32dsi28}, lower panel, we show that increasing
the neutron capture cross section of $^{32}$Si by
a factor of 100 (see discussion in \S \ref{sec: models_description})
does not change our results.
%In particular,
%%small effects are obtained by reducing the low $^{32}$Si MACS (which
%%is lower than 1 mb), whereas
%larger effects are obtained by using a higher
%$^{32}$Si MACS, making more efficient the production of heavier Si species
%during the neutron burst. On the other way,
By reducing the $^{32}$Si Maxwellian-averaged cross section (MACS)
of the same factor the $^{32}$Si/$^{28}$Si ratio increases by less than 10\%,
since the $^{32}$Si MACS adopted in our models is already lower than 1 mb,
behaving as a bottle-neck in the neutron capture flow feeding heavier Si species.
Note that at the temperatures of explosive He-burning
the half-life of $^{32}$Si can be reduced down to few days
\citep[e.g.,][]{oda:94}. However, the timescale of the explosive nucleosynthesis
is less than $\sim$ 0.3 secs, and the impact of the $^{32}$Si half-life
in the calculations is negligible.

We have shown that CCSN models can explain the large $^{32}$S-excess
measured in SiC C grains by the
radioactive decay of the unstable isotope $^{32}$Si after grain formation.
Furthermore, in SiC C grains most of the remaining S is coming
from contamination. We have identified two typical conditions where the correct
$^{32}$Si/$^{28}$Si ratio can be obtained, depending on the explosion temperature
and on the abundance of $^{28}$Si.

\section{Conclusions}
\label{sec: summary}

We have compared the isotopic signatures in presolar
SiC grains of type C with nucleosynthesis predictions
for CCSN ejecta exposed to different shock velocities.
We propose that the seemingly incompatible Si and S isotopic ratios
in these grains are explained by assuming that the $^{32}$S
excess observed today originates from radioactive $^{32}$Si
that condensed into the forming SiC grains, and
decayed into $^{32}$S at later stages.
Assuming that all the remaining S
%, including all $^{33,34}$S observed in C grains
is due to contamination,
we estimated the $^{32}$Si/$^{28}$Si ratio in the
%C-rich explosive He shell ejecta of the
parent CCSN ejecta, ranging from a few 10$^{-4}$ to a few 10$^{-3}$.
We propose this ratio to be a lower limit of its original value in the explosive
He shell layers, depending on the level of contamination
or mixing with more normal material for each C grain.
Such ratios can be produced for different shock velocities
and/or explosion energies.
Two typical conditions reproducing directly the observed
$^{32}$Si/$^{28}$Si ratios are: one with high temperature
and large $^{28}$Si abundance ($\sim1.6\times10^9  \mathrm{K}$
and $\sim 5\times10^{-2}$, respectively), and one with
temperature $\sim0.7-0.9\times10^9  \mathrm{K}$
and $^{28}$Si abundance  $\sim 5\times10^{-4}$.
In the first case the neutron density reaches a peak
of a few $10^{22} \mathrm{cm}^{-3}$ for
few 10$^{-5} \mathrm{sec}$, rapidly dropping to values
more typical of the $n$-process neutron-burst.
In the second case, the neutron density peak is on the order of
10$^{18-19} \mathrm{cm}^{-3}$.

In conclusion, C grains carry a record of the neutron density reached
in the explosive He shell of the CCSN where they formed.
We showed that the theoretical nuclear reactions
in the $^{32}$Si mass region have large uncertainties, but
our results are not significantly affected.
%, but for sure
%a better knowledge of these rates would help to better constrain
%stellar nucleosynthesis calculations.
%theoretical predictions in the $^{32}$Si mass
%region are strongly affected by uncertainties in nuclear properties,
%with neutron capture cross sections varying by almost two orders
%of magnitude according to the HF model used. This is the
%major source of uncertainty in the present results.
We conclude that C grains carry the signature of lower energy
ejecta compared to SiC X grains, showing positive $\delta$(Si)
values and a significant amount of
$^{32}$Si produced by neutron captures.

%Acknowledgements:
\acknowledgments
We thank the Anonymous Referee for the very careful review
of the paper, which significantly contributed to improve
the quality of the publication.
NuGrid acknowledges significant support from NSF grants PHY 02-16783
and PHY 09-22648 (Joint Institute for Nuclear Astrophysics, JINA) and
EU MIRG-CT-2006-046520. The continued work on codes and in disseminating
data is made possible through funding from STFC and EU-FP7-ERC-2012-St
Grant 306901 (RH, UK), and NSERC
Discovery grant (FH, Canada), and an Ambizione grant of the SNSF
(MP, Switzerland). MP TR RH and FKT also thank for support from EuroGENESIS.
NuGrid data is served by Canfar/CADC. EZ was supported by NASA grant NNX11AH14G.
MGB's research was carried out under the auspices of the National Nuclear
Security Administration of the U.S. Department of Energy at Los Alamos National
Laboratory under Contract No. DE-AC52-06NA25396.
PH thanks Ramanath Cowsik for his hospitality at the McDonnell Center for the Space Sciences at Washington University.
TR also acknowledges the support from the THEXO collaboration within
the EU 7th Framework Programme, the European Research Council,
and the Swiss NSF. RH also acknowledges support from the World
Premier International Research Center Initiative (WPI Initiative), MEXT, Japan,
and from European Research Council under the European Union Seventh
Framework Program (FP/2007-2013) / ERC Grant Agreement n. 306901.

%\bibliography{all,i-process}
\hyphenation{Post-Script Sprin-ger}

\clearpage

%%%%%%%%%%%%%%%%%%%%%%%%%

%\bibliography{astro}
%\bibliography{ms}

%%%%%%%%%%%%%%%%%
% tables
%%%%%%%%%%%%%%%%%

%\input{tab_network2.tex}
\newpage

% Figures

%%%%%%%%%%%%%%%%%%%%%%%%%

%%%%%%%%%%%%%%%%%%%%%%%%%%%%%%%%%%%%%%%%%%%%%%
%	PLOT HE CORE ISOTOPES EVOLUTION I
%%%%%%%%%%%%%%%%%%%%%%%%%%%%%%%%%%%%%%%%%%%%%%
%\clearpage
%\begin{figure}
%\centering
%\resizebox{6cm}{!}{\rotatebox{-90}{\includegraphics{he25z2m2he4c12o16_ms_v2.ps}}}
%\resizebox{6cm}{!}{\rotatebox{-90}{\includegraphics{he25z2m2ne22mg25mg26_ms2_v2.ps}}}
%\hfill
%\resizebox{6cm}{!}{\rotatebox{-90}{\includegraphics{he25z2m2sonly_ms_v2.ps}}}
%\resizebox{6cm}{!}{\rotatebox{-90}{\includegraphics{he25z2m2taunn_ms_v2.ps}}}
%\hfill
%\resizebox{6cm}{!}{\rotatebox{-90}{\includegraphics{he25z2m2cuzn_ms.ps}}}
%\resizebox{6cm}{!}{\rotatebox{-90}{\includegraphics{he25z2m2gage_ms.ps}}}
%  \caption{.}
%\label{}
%\end{figure}

\end{document}